\documentclass{article}

\usepackage{fullpage}
\usepackage{amsmath,amstext,amssymb}
\usepackage{url}
\usepackage{hyperref}

\def\K{{\sf{K}}}
\def\R{{\sf{R}}}

\newcommand{\ZZ}{{\mathbb Z}}

\begin{document}

\title{\bf \large {DIFFERENTIATION OF KALTOFEN'S \\
DIVISION-FREE DETERMINANT ALGORITHM}\\
\normalsize Abstract} 

\author{\normalsize Gilles Villard\\
\footnotesize  CNRS, Universit\'e de Lyon\\
\footnotesize Laboratoire LIP, CNRS-ENSL-INRIA-UCBL \\[-0.1cm]
\footnotesize 46, All\'ee d'Italie, 69364 Lyon Cedex 07, France\\[-0.1cm]
\footnotesize {\tt {http://perso.ens-lyon.fr/gilles.villard}}
}

\date{}

\maketitle

\begin{center}
\begin{minipage}{12cm}
{\small
Kaltofen has proposed a new approach in~\cite{Kal92} for computing matrix determinants.
The algorithm is based on a baby steps/giant steps construction of Krylov subspaces,
and computes the determinant as the constant term of a characteristic polynomial. 
For matrices over an abstract field and by the results of Baur and Strassen~\cite{BaSt82}, 
 the 
determinant algorithm, actually a straight-line program, 
leads to an algorithm with the same complexity for computing the adjoint of a matrix~\cite{Kal92}.
However, the latter is obtained by the reverse mode   
of automatic differentiation and somehow is not ``explicit''.
We study this adjoint algorithm, show how it can be implemented (without resorting to an automatic transformation),
and demonstrate its use on polynomial matrices.}
\end{minipage}
\end{center}

\vspace*{0.6cm}

Kaltofen has proposed in~\cite{Kal92} a new approach for computing matrix determinants.
This approach has brought breakthrough ideas for improving the complexity estimate for the problem of 
computing the determinant without divisions over an abstract ring~\cite{Kal92,Kavi04-2}. 
The same ideas also lead to the currently 
best known bit complexity estimates for some problems on integer matrices such as the 
problem of computing the characteristic polynomial~\cite{Kavi04-2}.

We consider the straigth-line programs of~\cite{Kal92}  
for computing the determinant over abstract fields or rings (with or without divisions).  
Using the reverse mode of automatic differentiation
(see~\cite{Lin70,Lin76,OWB71}), 
a straight-line program for computing the determinant 
of a matrix $A$ can be (automatically) transformed 
into a program for computing the adjoint matrix $A ^{*}$ of
$A$~\cite{BaSt82}
(see the application in~\cite[\S1.2]{Kal92} and~\cite[Theorem\,5.1]{Kavi04-2}). Since
the latter program is derived by an automatic process, few is known
about the way it computes the adjoint. 
The only available information seems to be the determinant program itself and the  
knowledge we have on the differentiation process. 
In this paper we study the adjoint programs that would be automatically generated by differentiation from  
Kaltofen's determinant programs. We show how they can be implemented with and without divisions, and 
study their behaviour on univariate polynomial matrices. 

Our motivation for studying the differentiation and resulting adjoint algorithms is the  
importance of the determinant approach of~\cite{Kal92,Kavi04-2} for various complexity estimates. 
Recent advances around the determinant of polynomial or integer matrices~\cite{GEV00,Kavi04-2,Sto03,Sto05},
and the adjoint of a univariate polynomial matrix in the generic case~\cite{JeVi06}, 
also justify the study of the general adjoint problem. 



\section{Kaltofen's determinant algorithm}

Let $\K$ be a commutative field. We consider $A \in \K ^{n \times n}$,
$u \in \K ^{ n \times 1}$, and $v \in \K ^{n \times 1}$. Kaltofen's approach extends
the Krylov-based methods of~\cite{Wie86,KaPa91,KaSa91}. 
We introduce the Hankel matrix $H= (uA^{i+j-2}v)_{ij} \in \K ^{n \times n}$,
and let $h_k=uA^kv$ for $0\leq k \leq 2n-1$.
We assume that $H$ is non-singular.
In the applications the latter  
is ensured either by construction of $A,u$, and
$v$~\cite{Kal92,Kavi04-2}, 
or by randomization (see~\cite{Kavi04-2} and references therein).

With baby steps/giant steps parameters $r=\lceil 2n/s \rceil$ and $s=\lceil
\sqrt{n}\rceil$ ($rs \geq 2n$) we consider the following
algorithm (the algorithm without divisions will be described in 
Section~\ref{sec:nodiv}). 

\begin{description}
\item Algorithm {\sc Det}~\cite{Kal92}
\item {\sc step 1.}  For $i=0,1,\ldots,r-1$ Do $v_i := A^i v$;\\[-0.6cm]
\item {\sc step 2.}  $B=A^r$; \\[-0.6cm]
\item {\sc step 3.}  For $j=0,1,\ldots,s-1$ Do $u_j := u B^j$;\\[-0.6cm]
\item {\sc step 4.}  For $i=0,1,\ldots,r-1$ Do \\
\hspace*{0.64cm}For $j=0,1,\ldots,s-1$ Do $h_{i+jr}:=u_jv_i$;  \\[-0.6cm]
\item {\sc step 5.} Compute the minimum polynomial $f(\lambda)$  of the sequence $\{h_k\}_{0\leq k \leq 2n-1}$;\\[-0.6cm]
\item Return $f(0)$. 
\end{description}


\section{The ajoint algorithm}

The determinant of $A$ is a polynomial in
$\K[a_{11},\ldots,a_{ij},\ldots,a_{nn}]$ of the entries of $A$. 
If we denote the adjoint matrix by $A ^{*}$ such that 
$A A ^{*}= A ^{*} A = (\det A) I$, then the entries of $A ^{*}$ satisfy~\cite{BaSt82}: 
\begin{equation} \label{eq:partial}
  a_{j,i}^{*}=\frac{\partial \Delta}{\partial a_{i,j}}, 1\leq i,j
  \leq n.
\end{equation}

The reverse mode of automatic differentiation (see~\cite{BaSt82,Lin70,Lin76,OWB71})
 allows to transform a
program which computes $\Delta$ into a program which computes all
the partial derivatives in~(\ref{eq:partial}).
We apply the transformation process to Algorithm~{\sc Det}.\\

The flow of computation for the adjoint is reversed compared to the
flow of Algorithm~{\sc Det}.
Hence we start with the differentiation of {\sc Step}~5.
Consider the $n \times n$ Hankel matrices $H=(uA^{i+j-2}v)_{ij}$
and $H_A=(uA^{i+j-1}v)_{ij}$. Then the determinant $f(0)$ is computed 
as 
$$
\Delta = (\det H_A)/(\det H). 
$$
Viewing $\Delta$ as a function $\Delta_5$ of the $h_k$'s, we show that 
\begin{equation} \label{eq:diff5}
\frac{\partial \Delta_5}{\partial h_k} = (\varphi_{k-1}(H_A ^{-1}) - \varphi_k(H
^{-1})) \Delta
\end{equation}
where for a matrix $M=(m_{ij})$ we define $\varphi_k(M)=
0+\sum_{i+j-2=k} m_{ij}$ for $1 \leq k \leq 2n-1$. 
Identity~(\ref{eq:diff5}) gives the first step of the adjoint
algorithm. Over an abstract field, and using intermediate data from
Algorithm~{\sc Det}, its costs 
is essentially the cost of a Hankel matrix inversion.  \\

For differentiating {\sc Step~4}, $\Delta$ is seen as a function
$\Delta _4$ of
the $v_i$'s and $u_j$'s. 
The entries of $v_i$ are involved in the computation of 
the $s$ scalars $h_i, h_{i+r}, \ldots, h_{i+(s-1)r}$.
The entries of $u_j$ are used for computing 
the $r$ scalars $h_{jr},h_{1+jr}, \ldots, h_{(r-1)+jr}$. 
Let $\partial v_i$ be the $1\times n$ vector, respectively the $n\times 1$
vector $\partial u_j$, whose entries are the
derivatives of $\Delta _4$ with respect to the entries of $v_i$,
respectively $u_j$. We show that 
\begin{equation} \label{eq:diff4a}
\left[\begin{array}{c}
~~~~\partial v_0~~~~\\
\partial v_1\\
\vdots\\
\partial v_{r-1}
\end{array}
\right]
= H^{v}   
\left[\begin{array}{c}
~~~~u_0~~~~\\
u_1\\
\vdots\\
u_{s-1}
\end{array}
\right]
\end{equation}
and 
\begin{equation} \label{eq:diff4b}
\left[\begin{array}{cccc}
\partial u_0, \partial u_1, \ldots \partial u_{s-1}
\end{array}
\right]
=  
\left[\begin{array}{cccc}
v_0, v_1, \ldots v_{r-1}
\end{array}
\right] H^{u} 
\end{equation}
where $H^{v}$ and  $H^{u}$ are 
$r \times s$ matrices whose entries are selected
$\partial \Delta _5 / \partial h_k$'s.   
Identities~(\ref{eq:diff4a}) and~(\ref{eq:diff4b}) give the second step of the adjoint
algorithm. Its costs 
is essentially the cost of two $n \times \sqrt{n}$ by  $\sqrt{n} \times \sqrt{n}$ 
(unstructured) matrix products.

Note that~(\ref{eq:diff5}),~(\ref{eq:diff4a}) and~(\ref{eq:diff4b})
somehow call to mind the matrix factorizations~\cite[(3.5)]{Ebe97}
(our objectives are similar to Eberly's ones) 
and~\cite[(3.1)]{EGGSV07}.\\

Steps~3-1 of {\sc Det} may then be differentiated. For  
differentiating {\sc Step 3} we recursively compute an $n\times n$ matrix $\partial B$ from the
$\delta u_j$'s. The matrix 
$\partial B$ gives the derivatives of $\Delta _3$ (the determinant seen 
as a function of $B$ and the $v_i$'s) with respect to the entries of
$B$.

For {\sc Step 2} we recursively compute from $\delta B$ an $n \times n$ matrix
$\delta A$ that gives the derivatives of $\Delta _2$ (the determinant seen 
as a function of $v_i$'s). 

Then the differentiation of {\sc Step 1} computes from $\delta A$
and the $\delta v_i$'s an update of $\delta A$
that gives the derivatives of $\Delta _1 =
\Delta$. From~(\ref{eq:partial}) we know that $A ^{*} = (\delta A)
^T$. \\

The recursive process for differentiating {\sc Step}~3 to {\sc
  Step}~1 may be written in terms of the differentiation of the basic operation 
(or its transposed operation) 
\begin{equation} \label{eq:pq}
q := p \times M  
\end{equation}
where $p$ and $q$ are row vectors of dimension $n$ and $M$ is an
$n\times n$ matrix. 
We assume at this point (recursive process) that column vectors 
$\delta p$ and $\delta q$ of derivatives with respect to the entries
of $p$ and $q$ are available. We also assume that an $n \times n$ 
matrix $\delta M$ that gives the derivatives with respect to the
$m_{ij}$'s has been computed. 
We show that differentiating~(\ref{eq:pq}) amounts to updating 
$\delta p$ and $\delta M$ as follows: 
\begin{equation} \label{eq:diffpq}
\left\{\begin{array}{l}
\delta p := \delta p + M \times \delta q,\\
\delta M := \delta M + p ^T \times (\delta q)^T.
\end{array} \right.
\end{equation}
We see that the complexity is essentially preserved
between~(\ref{eq:pq}) and~(\ref{eq:diffpq}) and corresponds 
to a matrix by vector product. 
In particular, if {\sc Step 2} of Algorithm {\sc Det} is implemented in $O(\log r)$
matrix products, then  
{\sc Step 2} differentiation 
will cost $O(n^3 \log r)$
operations (by decomposing the $O(n^3)$ matrix product).\\

Let us call {\sc Adjoint} the algorithm just described for computing
$A ^{*}$. 


\section{Application to computing the adjoint without divisions} \label{sec:nodiv}

Now let $A$ be an $n\times n$ matrix over an abstract ring $\R$.   
Kaltofen's method for computing the determinant of $A$ without divisions
applies Algorithm {\sc Det} on a well chosen univariate polynomial matrix $Z(z) =
C + z (A-C)$ where $C \in \ZZ ^{n \times n}$.
The choice of $C$ as well as a dedicated choice for the projections
$u$ and $v$ allow the use of  Strassen's 
general method of avoiding divisions~\cite{Str73,Kal92}.
The determinant is a polynomial $\Delta$ of degree $n$, 
the arithmetic operations in {\sc Det} are replaced by operations on
power series modulo $z^{n+1}$. 
Once the determinant of $Z(z)$ is computed, $(\det Z)(1) = \det (C +
1 \times (A-C))$ gives the determinant of $A$.

In {\sc Step}~1 and {\sc Step}~2 in Algorithm {\sc Det} applied to $Z(z)$ the 
matrix entries are actually polynomials of degree at most $\sqrt{n}$. 
This is a key point for reducing the overall
complexity estimate of the problem. Since the adjoint algorithm has
a reversed flow, this key point does not seem to be relevant 
for {\sc Adjoint}. For computing $\det A$ without divisions, Kaltofen's algorithm
goes through the computation of $\det Z(z)$. {\sc Adjoint} applied to
$Z(z)$ computes $A ^{*}$ but does not seem to compute $Z ^{*}(z)$ 
with the same complexity. In particular, differentiation of {\sc Step}~3
using~(\ref{eq:diffpq}) leads to products $A ^l (\delta B)^T$ 
that are more expensive over power series (one computes $A(z) ^l (\delta B(z))^T$)
than the initial computation in {\sc Det}
$A ^r$ ($A(z)^r$ on series).  
\\

For computing $A ^{*}$ without divisions only $Z ^{*}(1)$ needs to
be computed. We extend algorithm {\sc Adjoint} with input $Z(z)$    
by evaluating polynomials (truncated power series) partially.
With  a final evaluation at $z=1$ in mind, 
a polynomial $p(z) = p_0 + p_1 z + \ldots + p_{n-1} z^{n-1} + p_n z^n$ 
may typically be replaced by $(p_0 + p_1  + \ldots + p_m) + p_{m+1}
x^{m+1} + \ldots + p_{n-1} z^{n-1} + p_n z^n$ as soon as any subsequent
use of $p(z)$ will not require its coefficients of degree less than $m$.


\section{Fast matrix product and application to polynomial matrices}

We show how to integrate asymptotically fast matrix products in Algorithm {\sc Ajoint}. 
On univariate polynomial matrices $A(z)$ with power series operations modulo $z^n$, 
Algorithm {\sc Adjoint} leads to intermediary square matrix products where one of the operand   
has a degree much smaller than the other. 
In this case we show how to use fast rectangular matrix products~\cite{CoWi90,HuPa97} for a (tiny) improvement of 
the complexity estimate of general polynomial matrix inversion.

\section*{Concluding remarks}

Our understanding of the differentiation of Kaltofen's determinant algorithm has to be improved. 
We have proposed an implementation whose mathematical 
explanation remains to be given. Our work also has to be generalized
to the block algorithm of~\cite{Kavi04-2}.\\

\noindent
{\bf Acknoledgements.} We thank Erich Kaltofen who has brought reference~\cite{OWB71} to our attention.


\bibliographystyle{plain}


\end{document}